\null
\magnification=\magstep1
\hsize=16.2truecm
\vsize=23.5truecm
\voffset=0\baselineskip
\parindent=1truecm
\tolerance=10000
\baselineskip=17pt
\voffset=0\baselineskip
\parskip=0cm
\par
\def\al{\alpha}
\def\be{\beta}
\def\ga{\gamma}

\def\d{\partial}

\def\de{\delta}
\def\e{\epsilon}
\def\De{\Delta}

\def\ga{\gamma}

\def\la{\lambda}
\def\La{\Lambda}
\def\k{\kappa}

\def\N{{\cal N}}

\def\oo{\over}

\def\si{\sigma}

\def\ta{\tau}
\def\~{\tilde}
\def\^{\hat}
 
\noindent
{\bf VACUUM KERR-SCHILD METRICS}
 
\noindent
{\bf GENERATED BY NONTWISTING CONGRUENCES}\footnote{$^\dagger$}{Research
supported by OTKA fund no. 1826 and by the Pro Renovanda Cultura
Hungariae Foundation}\hfil \vskip.05in \noindent
{\sl L\'aszl\'o \'A. Gergely\footnote{$^a)$}{{\it Department of
Theoretical Physics, J¢zsef Attila University, H-6720 Szeged,
Aradi v'rtan£k tere 1, Hungary}}
and Zolt\'an Perj\'es\footnote{$^b)$}{{\it Central Research
Institute
for Physics, H-1525 Budapest 114, P.O.Box 49, Hungary}} }
\parskip=0.5truecm
\vskip.05in
\noindent
{\bf ABSTRACT}
\midinsert
\baselineskip=12pt plus.1pt
   The Kerr-Schild pencil of metrics $\tilde g_{ab}=g_{ab}+V
l_al_b$, with $g_{ab}$ and $\tilde g_{ab}$ satisfying
the vacuum Einstein equations,
is investigated in the case when the null vector
$l$ has vanishing twist. This class of Kerr-Schild metrics
contains two solutions: the Kasner metric and a metric wich can be
obtained from the Kasner metric by a complex coordinate
transformation. Both are limiting cases of the
K\'ota-Perj\'es metrics. The base space-time is a pp-wave.
 
 
\endinsert
\baselineskip=17pt
\voffset=0\baselineskip
\par
 
\noindent
{\bf 1. INTRODUCTION}
 
      The Kerr-Schild map
$$\tilde g_{ab}=g_{ab}+V l_al_b    \eqno (0)$$
generates a pencil of space-time metrics $\tilde g_{ab}$ from the metric
$g_{ab}$ where $l$ is a null vector and $V$ a function. The
solution of the original problem, where a {\sl flat} parent
space-time is mapped to a vacuum
space-time, has been known for some time.
In the case of Kerr-Schild pencils with a non-twisting and
divergence-free $l$, this is due to Trautman$^1$. For the congruences
with nonvanishing divergence or twist,
the general solution is given by Kerr and Schild$^2$.
 
 Here we shall not impose any restriction on the parent metric $g$
apart from being Lorentzian and vacuum. Still, it follows from the
vacuum Einstein equations for the pencil (0) that the null vector
$l$ is tangent to geodesics.
The general solution of this problem was given recently
by us$^{3,4,5}$. However, the special case when $l$ is twist-free is
not {\sl per se} covered by the generic procedure, and hints only
were given in Ref. 5 as to how to treat the twist-free
fields.
The full details are given in this paper.
\vfill\eject

   In section 2 we state a theorem characterizing the
twist-free Kerr-Schild metrics by the value of the shear
parameter $\eta$. As in the generic case, the values
$sin\eta=0,\pm 1, \pm {1\oo\sqrt 2}$ of the shear parameter
are exceptional, and these metrics cannot be continuously extended
to other values of $\eta$. These exceptional cases are covered by
the treatment in Ref. 5, letting the twist vanish. However, the
generic treatment
in Ref. 5 cannot be extended to the fields with arbitrary
$\eta$. Sections 3 and 4 are devoted to these fields.
Nonetheless, the resulting metrics prove to be the twist-free
limits of the generic metrics$^5$.
 
\bigskip
\noindent
{\bf 2. SOLUTION OF THE $\Psi_1$ SYSTEM}
 
    The computations follow closely the pattern of the generic
case where the Newman-Penrose formalism$^6$ was used. The
tetrad vectors are denoted $l=D,\ n=\De,\ m=\de,$
and $\bar m=\bar\de$. The expressions
for the optical scalars, the curvature quantity $\Psi_0$, the
Kerr-Schild potential and the null tetrad vector $m$
in the twist-free case ($B=0$) become$^{5}$:
$$\eqalign{\rho=&-{1+cos\eta\oo2r}\ ,\quad
           \si=-{sin\eta\oo2r}\ ,\quad
           \Psi_0=-{sin\eta cos\eta\oo2r^2}\ ,\cr
           V=&{V_0\oo r^{cos\eta}}\ ,\quad
           m=r^{-{cos\eta+sin\eta+1\oo2}}(Q_{\bf
1}+ir^{sin\eta}Q_{\bf 2})\
.} \eqno(1)$$
where
$Q_{\bf 1,2}=Q_{\bf 1,2}^j {\d\oo\d x^j}$.
The functions $\eta,Q_{\bf 1}^j,
Q_{\bf 2}^j$  and $V_0$ do not depend on $r$,
the affine parameter of the (geodesic) integral
curves of $l=D={\d\oo\d r}$.
The tetrad has been uniquely fixed by choosing
$$\e=0\ ,\quad \pi=\al+\bar\be\ , \quad m^0=0 \ . \eqno(2)$$
The fields with $\eta=0$ are algebraically special.
 
      Our main theorem$^4$ can be stated in the twist-free case
again:
 
{\sl Theorem on twist-free vacuum-vacuum Kerr-Schild maps:}
Unless $sin\eta=0,\pm 1, \pm {1\oo\sqrt 2}$, the
Kerr-Schild potential is restricted by the relation $\de V=0$.
 
    The treatment of the metrics with $sin\eta=0,\pm 1, \pm {1\oo\sqrt 2}$
in Ref. 5 holds unchanged for curl-free fields. In what follows,
we will consider the case where the shear is not restricted.
The explicit $r$ dependence of $\de V$ is known by (1). Hence we
obtain:
$$\de V_0=\de\eta=0\ . \eqno(3)$$
  Though the forms of the spin coefficients (1) are simple under
the curl-free condition, the integration functions $V_0$ and
$\eta$ are not constant. (Unlike in the generic case
$\bar\rho-\rho\neq0$, it does not follow from the commutator
$[\bar\de,\de]$ that their $\De$ derivatives vanish.)
 
   The field equations are the Newman-Penrose equations
and the vacuum Einstein-equations for Kerr-Schild metrics$^4$.
The subsystem of equations containing $D$ derivatives of
spin coefficients $\al,\be,\ta,\pi$ and $\Psi_1$ is closed and has
been referred to
as the $\Psi_1$ system$^{3,4}$. The general solution for this
system was obtained as a finite series in the complex phase factor
$C={r^{cos\eta}-iB\oo r^{cos\eta}+iB}$. For the present nontwisting
case the function $B=0$, thus $C=1$. When we take the limit $B\to0$
for the four fundamental solutions of the $\Psi_1$ system, two of
them vanish because they are
proportional to $C-1$. The fundamental solutions can be multiplied by
arbitrary $r$-independent real functions. Thus we
divide by $B$ and apply the l'Hospital rule,
${C-1\oo B}\ \to\ -{2i\oo r^{cos\eta}}$ to obtain the
two missing fundamental solutions of the $\Psi_1$ system.
 The general solution is:
$$\eqalign{
\pi=&{E_1\pi^{(1)}\oo r^{cos\eta+sin\eta+1\oo2}}+
     {E_2\pi^{(2)}\oo r^{cos\eta-sin\eta+1\oo2}}+
     {E_3\pi^{(3)}\oo r^{3cos\eta+sin\eta+3\oo2}}+
     {E_4\pi^{(4)}\oo r^{3cos\eta-sin\eta+3\oo2}}\cr
\ta=&{E_1\ta^{(1)}\oo r^{cos\eta+sin\eta+1\oo2}}+
     {E_2\ta^{(2)}\oo r^{cos\eta-sin\eta+1\oo2}}+
     {E_3\ta^{(3)}\oo r^{3cos\eta+sin\eta+3\oo2}}+
     {E_4\ta^{(4)}\oo r^{3cos\eta-sin\eta+3\oo2}}\cr
\al=&{E_1\al^{(1)}\oo r^{cos\eta+sin\eta+1\oo2}}+
     {E_2\al^{(2)}\oo r^{cos\eta-sin\eta+1\oo2}}+
     {E_3\al^{(3)}\oo r^{3cos\eta+sin\eta+3\oo2}}+
     {E_4\al^{(4)}\oo r^{3cos\eta-sin\eta+3\oo2}}\cr
\Psi_1=&{E_1\Psi^{(1)}_1\oo r^{cos\eta+sin\eta+3\oo2}}+
        {E_2\Psi^{(2)}_1\oo r^{cos\eta-sin\eta+3\oo2}}+
        {E_3\Psi^{(3)}_1\oo r^{3cos\eta+sin\eta+5\oo2}}+
        {E_4\Psi^{(4)}_1\oo r^{3cos\eta-sin\eta+5\oo2}}}
\eqno(4)$$
where $E_1,...E_4$ are arbitrary functions independent of $r$
and the coefficients $\pi^{(1)},...,$
$\pi^{(4)},\ta^{(1)},...\ta^{(4)},\al^{(1)},...\al^{(4)},\Psi^{(1)}_1,...\Psi^{(4)}_1$
enlisted in Table 1 are functions of $\eta$ only.
 
      Once the affine-parameter dependence of the spin
coefficients $\rho,\si,\al,\be,\pi,\ta,\Psi_0,\Psi_1$ and of the
Kerr-Schild potential is known, the question naturally arises whether
these solutions are compatible with the rest of the equations? In
the following section, we will address this question.
 
\vfill\eject
\noindent
{\bf 3. THE REMAINING FIELD EQUATIONS}
 
      First we find from the commutator
$[\bar\de,\de]r=\mu-\bar\mu$ that
the spin coefficient $\mu$ is real. Then, writing the tetrad
vector $n$ in the form
$$n=\N \partial/\partial r +N^j \partial/\partial x^j  \eqno(5)$$
where $\N,N^j$ are unknown functions of all of the coordinates,
some $\De$ derivatives can be written as:
$$\eqalign{
\De\rho=&{1+cos\eta\oo2r^2}\N+{sin\eta\oo2r}\De\eta\ ,\qquad
\De\si={1+cos\eta\oo2r^2}\N+{sin\eta\oo2r}\De\eta\ ,\cr
\De\Psi_0=&{sin\eta cos\eta\oo r^3}\N
          -{cos^2\eta-sin^2\eta\oo2r^2}\De\eta\ ,\quad
\De ln V=\De ln V_0-{cos\eta\oo r}\N+sin\eta\ ln r\ \De\eta
\ .}\eqno(6)$$
 
      The equations (NP 4.2.l,q) of Ref. 6, the complex conjugate
of (NP 4.2.p), the fifth Bianchi relation (NP 4.5) and the last
Kerr-Schild equation$^{3,5}$ form a closed algebraic system in
the real variables $\mu,\N,\De\eta,\De ln V_0$ and the complex
variables $\la,\ga,\Psi_2$ and their complex conjugates:
$$\eqalign{
\rho\mu-\si\la-\Psi_2&=a_1 \cr
\si\mu+\rho\la+\si(\ga-3\bar\ga)
 +{1+cos\eta\oo2r^2}\N+{sin\eta\oo2r}\De\eta&=\bar a_2 \cr
\rho\mu+\si\la-\rho(\ga+\bar\ga)+\Psi_2
 +{sin\eta\oo2r^2}\N-{cos\eta\oo2r}\De\eta&=a_3\cr
\Psi_0\mu-4\Psi_0\ga-3\si\Psi_2
 +{sin\eta cos\eta\oo r^3}\N-{cos^2\eta-sin^2\eta\oo2r^2}\De\eta
     &=a_4 \cr
{\Psi_0\oo\si}\mu-2\rho(\ga+\bar\ga)+\Psi_2+\bar\Psi_2
 -\rho(\De ln V_0-{cos\eta\oo r}\N+sin\eta\ ln r\ \De\eta)
     &=a_5 }\eqno(7)$$
where the source terms $a_1,...a_5$ are given in Table 2, and
their $r$ dependence is known. This system corresponds to, but
does not hold as a limiting case, of Eqs. (5.2) of Ref. 5.
The unknown functions $\N,\De\eta$ and $\De ln V_0$ did not occur
in the generic treatment. Now we cannot get rid of $\N$ by use of
the commutator $[\de,\bar\de]r$ because $(\bar\rho-\rho)\N$
vanishes, and we cannot prove property (ii) (that the $\De$
derivatives vanish).
 
      The imaginary parts of the first four equations of (7)
yield:
$$\eqalign{
\la-\bar\la=&{-3\si^2(a_1-\bar a_1)+\si(a_4-\bar
    a_4)-\Psi_0(a_2-\bar a_2)\oo3\si^3+\rho\Psi_0}\cr
\ga-\bar\ga=&-{a_2-\bar a_2+\rho(\la-\bar\la)\oo4\si}\cr
\Psi_2-\bar\Psi_2=&-\si(\la-\bar\la)-(a_1-\bar a_1)\cr
C_1\equiv &a_1+a_3-\bar a_1-\bar a_3=0\ }\eqno(8)$$
The last relation is a constraint equation involving only
expressions with known $r$ dependence.
 
      With algebraic manipulations on the first four equations of (7)
one obtains:
$$\eqalign{
\mu=&{1+cos\eta\oo sin\eta}\la+f_1\cr
r\Psi_2=&-{cos\eta(1+cos\eta)\oo sin\eta}\la+f_2\cr
\ga+\bar\ga+{\N\oo r}=&{2(1+cos\eta)\oo sin\eta}\la+f_3\cr
\De\eta=&{2r(a_1+a_2)+(1+cos\eta)(2f_1-f_3)\oo sin\eta}\equiv r\
C_2+r^2\ C_3 } \eqno(9)$$
where $f_1,f_2,f_3$ enlisted in Table 3 were formed from $a_1,...a_4$.
Note that the commutator $[\De,D]\eta$ implies that $D\eta$ does
not depend on $r$, so $\De\eta=0$ and $C_2=C_3=0$ are constraints.
 
      From the equations (NP 4.2.g) the $r$ dependence of $\la$
can be integrated:
$$D\bigl(r^{1+cos\eta}\la\bigr)=r^{1+cos\eta}(\si
f_1+f_4)\eqno(10)$$
where the term $f_4$ with known $r$ dependence is given in Table 3.
After the elimination of $\la$ from (10) and (NP 4.2.h), a fourth
constraint $C_4=0$ arises. The solution of all these
constraints allows one to express the $Q_{\bf 1,2}$ derivatives of
$E_1,E_2$ in terms of $E_1,E_2$, while $E_3,E_4$ is found to
vanish (Table 4).
 
      Finally, from $[\De,D]r$ one gets $\bigl(\ga+\bar\ga+{\N\oo
r}\bigr)= rD\bigl(-{\N\oo r}\bigr)$, and with a second integration
one finds the $r$ dependence of $\N$.
 
      Using all these results in the last relation of (7), we
express $\De ln V_0$, which does not depend on $r$ as follows
from the commutator $[\De,D]ln V_0$. This condition can be
fulfilled only if:
$$E_1=E_2=E_3=E_4=0\ .\eqno(11) $$
We recover the same result as in the general case: only the
trivial solution of the $\Psi_1$ system is compatible with
the rest of the equations. We then have
$$\eqalign{
&\k=\e=\pi=\ta=\al=\Psi_1=\mu=\la=\Psi_2=\De\eta=0\cr
&\ga=-{H\oo2}\ ,\qquad \N=Hr\ ,\qquad \De ln
V_0=(cos\eta+2)H \ ,}\eqno(12)$$
where $H$ is an integration function restricted by the
conditions $Q_{\bf 1}H=Q_{\bf 2}H=0$ following from the commutator
$[\de,D]ln V_0$.
 
      We complete the process of determining the $r$ dependence of
the spin coefficients by observing that
$$\nu=\Psi_3=\Psi_4=0\ ,\eqno(13)$$
from $[\de,D]r$ and (NP 4.2.i,j) respectively. We conclude
that the base space must be of type N.
 
\bigskip
\bigskip
\noindent
{\bf 4. THE METRICS}
 
      The commutators among $Q_{\bf 1},Q_{\bf 2},N=N^j{\d\oo\d r}$
and $D$, namely
$$\eqalign{&[N,D]=[Q_{\bf 1},D]=[Q_{\bf 2},D]=[Q_{\bf 1},Q_{\bf 2}]=0\cr
&[Q_{\bf 1},N]=-{cos\eta+sin\eta+1\oo2}HQ_{\bf 1}\ ,\qquad
 [Q_{\bf 2},N]=-{cos\eta-sin\eta+1\oo2}HQ_{\bf 2}
}\eqno(14)$$
allow one to fix the coordinates $(r,x,y,u)$ in the following
way:
$$\eqalign{D=&{\d\oo\d r}\ ,\qquad Q_{\bf 1}={\d\oo\d x}\ ,\qquad
Q_{\bf 2}={\d\oo\d y}\ ,\cr
N=&-{cos\eta+sin\eta+1\oo2}H(u){\d\oo\d x}
-{cos\eta-sin\eta+1\oo2}H(u){\d\oo\d y}+F(u){\d\oo\d u}}\ ,
\eqno(15)$$
where $F(u)$ is an arbitrary function of $u$. The tetrad vectors
in the chosen coordinates are:
$$\eqalign{
&l^a=\de^a_1\qquad\qquad
m^a=r^{-{cos\eta+sin\eta+1\oo2}}\de^a_2
    +r^{-{cos\eta-sin\eta+1\oo2}}\de^a_3\cr
&n^a=ur\de^a_1-{cos\eta+sin\eta+1\oo2}ux\de^a_2
    -{cos\eta-sin\eta+1\oo2}uy\de^a_3+F(u)\de^a_4
}\eqno(16)$$
 
      The Kerr-Schild potential is found by integration of
$V_0$ from (12):
$$V=\La {e^{(cos\eta+2)\int{udu\oo F(u)}}\oo r^{cos\eta}}\ .\eqno(17)$$
where $\La$ is an arbitrary parameter.
 
      We are now able to write down the base space metric
using the completeness relation of the tetrad (16). With (17), the
Kerr-Schild pencil has the form
 $$\eqalign{
ds^2=&-{r^{cos\eta+sin\eta+1}\oo2}
            \left[dx+{(cos\eta+sin\eta+1)xu\oo2F(u)}du\right]^2\cr
     &-{r^{cos\eta-sin\eta+1}\oo2}
            \left[dy+{(cos\eta-sin\eta+1)yu\oo2F(u)}du\right]^2\cr
&-{1\oo2ru}\left[dr-{2ru\oo F(u)}du\right]^2+{dr^2\oo2ru}\cr
d\tilde s^2=&ds^2+\La{e^{(cos\eta+2)\int{udu\oo F(u)}}\oo
r^{cos\eta}}{du^2\oo F(u)^2} }\eqno(18)$$
 
     The curvature components are:
$$\eqalign{
&\tilde\Psi_0=-{cos\eta sin\eta\oo2r^2}\cr
&\tilde\Psi_1=\tilde\Psi_3=0\cr
&\tilde\Psi_2=\La{cos\eta(cos\eta+1)e^{(cos\eta+2)\int{udu\oo F(u)}}
        \oo4r^{cos\eta+2}}\cr
&\tilde\Psi_4=-\La^2{cos\eta sin\eta e^{2(cos\eta+2)\int{udu\oo F(u)}}
        \oo8r^{2cos\eta+2}}
\ .}\eqno(19)$$
We conclude that the Kerr-Schild space-time is of type I in the
Petrov classification.
 
      We may transform these metrics to a
simpler form by the coordinate transformations
$$t=e^{\int{udu\oo F(u)}}\ ,\quad
v={1\oo\sqrt2}xt^{cos\eta+sin\eta+1\oo2}\ ,\quad
w={1\oo\sqrt2}yt^{cos\eta-sin\eta+1\oo2}\ ,\quad
p=rt\ .\eqno(20)$$
Noting that $t$ is function only of $u$, we define a new
coordinate $q$ in place of $t$ by $dq={dt\oo u}$. The old
coordinates $(r,x,y,u)$ can be expressed in terms of the new
ones $(v,w,p,q)$ provided that the first relation of (20)
is invertible. The base and Kerr-Schild metrics in these new
coordinates are:
$$\eqalign{
ds^2=&-p^{cos\eta+sin\eta+1}dv^2
      -p^{cos\eta-sin\eta+1}dw^2
      +2dp dq\cr
d\tilde
s^2=&ds^2+\La p^{-cos\eta}dq^2
\ .}\eqno(21)$$
The parent space metric has been obtained by Bilge$^7$.
There has been some controversy in the literature (cf. Kupeli$^8$
and Bilge$^7$) whether or not this pp-wave is unique in the
considered class of Kerr-Schild maps.
Here we succeeded in demonstrating the uniqueness.
 
      The vectors ${\d\oo\d v}$, ${\d\oo\d w}$ and ${\d\oo\d q}$ are
(commuting) Killing vectors of both metrics, moreover the last of
them is covariantly constant.
In the shear-free limit $sin\eta\to0$ both metrics admit a
fourth Killing vector $v{\d\oo\d w}-w{\d\oo\d v}$, thus they are
plane symmetric metrics (13.9) of Ref. 9.
 
      The metric $d\tilde s^2$ in (21) can be put into a more
familiar form as follows,
\item{(a)} in case $\La<0$ we introduce the new coordinates
$(T,X,Y,Z)$:
$$\eqalign{
T=&{(-\La)^{-1\oo2}\oo{cos\eta\oo2}+1}p^{{cos\eta\oo2}+1}\cr
X=&\left[(-\La)^{1\oo2}\left({cos\eta\oo2}+1\right)\right]^
{cos\eta+sin\eta+1\oo2}v\cr
Y=&\left[(-\La)^{1\oo2}\left({cos\eta\oo2}+1\right)\right]^
{cos\eta-sin\eta+1\oo2}w\cr
Z=&\left[(-\La)^{1\oo2}\left({cos\eta\oo2}+1\right)\right]^
{-cos\eta\oo2}
\left[(-\La)^{1\oo2}q-(-\La)^{-1\oo2}{p^{1+cos\eta}\oo1+cos\eta}\right]
\ .}\eqno(22)$$
The Kerr-Schild metric is the Kasner metric$^{10}$:
$$d\tilde
s^2=dT^2-T^{2p_1}dX^2-T^{2p_2}dY^2-T^{2p_3}dZ^2 \
,\eqno(23)$$
where the powers
$$p_1={cos\eta+sin\eta+1\oo cos\eta+2}\ ,\quad
p_2={cos\eta-sin\eta+1\oo cos\eta+2}\ ,\quad
p_3={-cos\eta\oo cos\eta+2} \eqno(24)$$
satisfy the required relations $p_1+p_2+p_3=1$ and
$(p_1)^2+(p_2)^2+(p_3)^2=1$.
 
\item{(b)} in case $\La>0$ the new coordinates
$(T,X,Y,Z)$ are introduced in the following manner:
$$\eqalign{
T=&{(\La)^{-1\oo2}\oo{cos\eta\oo2}+1}p^{{cos\eta\oo2}+1}\cr
X=&\left[(\La)^{1\oo2}\left({cos\eta\oo2}+1\right)\right]^
{cos\eta+sin\eta+1\oo2}v\cr
Y=&\left[(\La)^{1\oo2}\left({cos\eta\oo2}+1\right)\right]^
{cos\eta-sin\eta+1\oo2}w\cr
Z=&\left[(\La)^{1\oo2}\left({cos\eta\oo2}+1\right)\right]^
{-cos\eta\oo2}
\left[(\La)^{1\oo2}q+(\La)^{-1\oo2}{p^{1+cos\eta}\oo1+cos\eta}\right]
\ .}\eqno(25)$$
The Kerr-Schild metric is a sign-flipped version of the Kasner metric
as described by McIntosh$^{11}$:
$$d\tilde
s^2=-dT^2-T^{2p_1}dX^2-T^{2p_2}dY^2+T^{2p_3}dZ^2 \
.\eqno(26)$$
This metric arises by way of a complex coordinate
transformation on the Kasner metric.
 
\bigskip
\noindent
{\bf 5. CONCLUDING REMARKS}
 
      Both solutions (23) and (26) are the special cases of
the K\'ota-Perj\'es metric (5.11) of Ref. 12 for $B=0$.
 
 \item{\it Lemma.}{The Kasner and sign-flipped Kasner metrics (24)
and (26), respectively, are the exhaustive members of the class of
vacuum Kerr-Schild metrics generated by nontwisting congruences}.
 
      Bilge$^7$ has considered vacuum metrics with a
non-twisting geodesic congruence and subject to the conditions
that $\sigma=a\rho$ and $Da=0$. He has proven that $a$ is a
constant. The relation of our work with Bilge's
can be established by noting that
his $a={\sin\eta\oo1+\cos\eta}$.
Bilge's solution involves the exceptional values
$\sin\eta=0,\ \pm 1,\pm4/5$
which are different from the ones given in our Theorem.
Apparently, the reason is that
the Kerr-Schild property has not been imposed
on the metric $\tilde g_{ab}$ by Bilge.

\vfill\eject\noindent
 
{\ \ }
\medskip
{\bf REFERENCES}
\frenchspacing
\smallskip
 
\item{[1]}  Trautman, A., in {\it Recent Developments in
General Relativity}, Pergamon Press ,459 (1962)
\item{[2]} Kerr, R. P.  and Schild, A., {\it Atti Del
Convegno Sulla Relativit  Generale: Problemi Dell' Energia e Onde
Gravitazionali (Anniversary Volume, Forth Centenary of Galileo's
Birth)}, G. Barb'ra, Ed. (Firenze, 1965), p. 173
\item{[3]} Gergely, \'A. L., Perj\'es Z., Physics Letters A {\bf
181}, 345 (1993)
\item{[4]} Gergely, \'A. L., Perj\'es Z., J. Math. Phys. {\bf
35}, 2438 (1994)
\item{[5]} Gergely, \'A. L., Perj\'es Z., J. Math. Phys. {\bf
35}, 2448 (1994)
\item{[6]} Newman, E. and Penrose, R., J. Math. Phys. {\bf3},
566 (1962)
\item{[7]} Bilge, A. H., Class. Quantum Grav. {\bf6}, 823 (1989)
\item{[8]} Kupeli, A. H., Class. Quantum Grav. {\bf 5}, 401 (1988)
\item{[9]} Kramer, D., {\it et al.}: {\it Exact Solutions of
Einstein Field Equations}, Cambridge Univ. Press (1980)
\item{[10]} Kasner, E., Amer. J. Math. {\bf 43}, 217 (1921)
\item{[11]} McIntosh, C. B. G., in {\it Relativity Today},
Z. Perj\'es, Ed., Nova Science Publishers, New York (1992), p. 147
\item{[12]} K\'ota, J. and Perj\'es, Z., J. Math. Phys. {\bf13}, 1695
(1972)
 
\vfill\eject\noindent
 
$$\eqalign{
\pi^{(1)}=&-2(3cos\eta+4sin\eta)(cos\eta-1)cos\eta\cr
\pi^{(2)}=&-2(4cos^2\eta+3cos\eta
       sin\eta-3cos\eta+4sin\eta-4)(cos\eta-1)\cr
\pi^{(3)}=&4\left[(sin\eta+3)cos\eta+(sin\eta+1)+(sin\eta-1)cos^2\eta-3cos^3\eta\right]\cr
\pi^{(4)}=&-4\left[(sin\eta+1)cos^2\eta+(sin\eta-1)+(sin\eta-3)cos\eta+3cos^3\eta\right]\cr\cr
\ta^{(1)}=&2(cos^2\eta-2cos\eta
       sin\eta+2cos\eta+2sin\eta-2)(cos\eta-1)\cr
\ta^{(2)}=&2(2cos\eta-sin\eta-1)(cos\eta-1)cos\eta\cr
\ta^{(3)}=&\pi^{(3)}\cr
\ta^{(4)}=&-\pi^{(4)}\cr\cr
\al^{(1)}=&\left[4(sin\eta+1)-(sin\eta-3)cos\eta-7cos^2\eta\right]cos\eta\cr
\al^{(2)}=&2(cos\eta+4)(sin\eta-1)-(7sin\eta-11)cos^2\eta-cos^3\eta\cr
\al^{(3)}=&{\pi^{(3)}\oo2}\cr
\al^{(4)}=&{\pi^{(4)}\oo2}\cr\cr
\Psi^{(1)}_1=&(cos^2\eta-7cos\eta
              sin\eta-3cos\eta-4sin\eta-4)(cos\eta-1)cos\eta\cr
\Psi^{(2)}_1=&(7cos^3\eta-cos^2\eta sin\eta+5cos^2\eta
              +10cos\eta sin\eta-10cos\eta+8sin\eta-8)(cos\eta-1)\cr
\Psi^{(3)}_1=&2(2cos^2\eta-4cos\eta sin\eta+cos\eta
             +sin\eta-3)(cos\eta+1)cos\eta\cr
\Psi^{(4)}_1=&-2(2cos^2\eta+4cos\eta sin\eta+cos\eta-sin\eta-3)
             (cos\eta+1)cos\eta\ }$$
 
{\it Table 1. The coefficients in the general solution of
the $\Psi_1$ system}
 
\bigskip
 
$$\eqalign{
a_1=&\de\al-\bar\de\be-\al\bar\al-\be\bar\be+2\al\be\ ,\qquad
a_2=\de\ta-\ta(\ta+\be-\bar\al)\ ,\cr
a_3=&\bar\de\ta+\ta(\bar\be-\al-\bar\ta)\ ,\qquad
a_4=\de\Psi_1-\Psi_1(4\ta+2\be)\cr
a_5=&\de(\bar\ta-\pi)+\bar\de(\ta-\bar\pi)-6\pi\bar\pi+2\pi\bar\al
+2\bar\pi\al-2\ta\bar\ta
-2\al\ta-2\bar\al\bar\ta+3\ta\pi+3\bar\ta\bar\pi }  $$
 
{\it Table 2. The source terms of eq. (7)}
 
\vfill\eject\noindent
 
\bigskip
 
$$\eqalign{
f_1&=-{2ra_1\oo 1+c}-{2b_1\oo s(1+c)}-{2[sb_2+2cr(a_1+a_2)]\oo
    (1+c)^2}\cr
f_2&={b_1\oo s}+{[sb_2+2cr(a_1+a_2)]\oo 1+c}\cr
f_3&=-2ra_1+{2b_1\oo s}-{[sb_2+2cr(a_1+a_2)]\oo 1+c}\cr
f_4&=\bar\de\pi+\bar\si \mu+\pi^2+(\al-\bar\be)\pi}$$
$$\eqalign{where: s&=sin\eta\ ,\qquad\ c=cos\eta\ ,\cr
b_1&=2r^2a_4-2r(c\bar a_2+sa_3)-4sc(\ga-\bar\ga)\ ,\qquad
b_2=2r\bar a_2+2s(\ga-\bar\ga)}$$
 
{\it Table 3. The inhomogeneous terms in eq. (9,10)}
 
\bigskip
\parindent=7pt
 
$$\eqalign{
(30&c^5+40c^4s-12c^4-16c^3s+2c^3-59c^2s+4c^2+18cs-34c-6s+6)\
Q_{\bf 1}(E_1)=\cr
&2E_1^2(110c^8-20c^7s-174c^7+168c^6s-184c^6-347c^5s+546c^5+127c^4s\cr
&\qquad-290c^4+296c^3s-220c^3-352c^2s+340c^2+152cs-152c-24s+24)}$$
 
$$(2c+s-1)\ Q_{\bf 2}(E_1)=-[Q_{\bf 1}(E_2)(c+2s-2)
+(12cs^2-12cs+40s^4-40s^3-24s^2+24s)E_1E_2]$$
 
$$\eqalign{&(7c^2-24cs-40c+30s+34)\ Q_{\bf 1}(E_2)=\cr
&2E_1E_2(82c^5+76c^4s-9c^4-237c^3s-289c^3+136c^2s+244c^2\cr
&\qquad +36cs-36c-8s+8)}$$
 
$$\eqalign{
(8c^4&-6c^3s+2c^3-4c^2s-5c^2+6cs+s-1)\ Q_{\bf 2}(E_2)=\cr
&2c^2E_2^2(10c^5-20c^4s-18c^4+16c^3s+8c^3+25c^2s\cr
&\qquad+10c^2-25cs-18c+4s+8) }$$
 
$$E_3=E_4=0$$
 
{\it Table 4. The solution of constraints $C_1-C_4$ for
$Q_{\bf i}(E_k)$, $E_3$ and $E_4$}
 
\end